\documentclass{KapProc} 

\usepackage{epsfig}
\usepackage{graphicx}
\normallatexbib

\begin{document}

\articletitle{Diagrammatic Theory of \\the Anderson impurity Model\\ 
  with finite Coulomb Interaction} 
\author{K. Haule$^{a,b}$, S.
  Kirchner$^b$, H. Kroha$^b$, P. W\"olfle$^b$}
\affil{$^a$ Institute Josef Stefan, Department of Theoretical Physics (F1)\\
\hspace*{0.3cm}SI-1000 Ljubljana, Slovenia\\
  $^b$ Institut f\"ur Theorie der Kondensierten Materie, Universit\"at
  Karlsruhe\\ \hspace*{0.3cm}D-76128 Karlsruhe, Germany}




\chaptitlerunninghead{Diagrammatic theory of the Anderson model with finite $U$}


\begin{abstract}
  We developed a self-consistent conserving pseudo particle
  approximation for the Anderson impurity model with finite Coulomb
  interaction, derivable from a Luttinger Ward functional. It contains
  an infinite series of skeleton diagrams built out of fully
  renormalized Green's functions. The choice of diagrams is motivated
  by the Schrieffer Wolff transformation which shows that singly and
  doubly occupied states should appear in all bare diagrams
  symmetrically.  Our numerical results for $T_K$ are in excellent
  agreement with the exact values known from the Bethe ansatz
  solution. The low energy physics of non-Fermi liquid Anderson
  impurity systems is correctly described while the present
  approximation fails to describe Fermi liquid systems, since some
  important coherent spin flip and charge transfer processes are not
  yet included. It is believed that CTMA (Conserving T-matrix
  approximation) diagrams will recover also Fermi liquid behavior for
  Anderson models with finite Coulomb interaction as they do for
  infinite Coulomb interaction.
\end{abstract}

\section{Introduction}
Anderson impurity models have been of considerable interest recently
for studying the transport in mesoscopic structures \cite{review} as
well as for describing strong correlations on the lattice in the limit
of large dimensions \cite{Vollhardt}. In both cases the doubly
occupied impurity site, which is characterized by an additional energy
cost $U$, is essential for the correct physical description.

In terms of pseudofermion ($f_\sigma$) and slave boson operators ($a$,
$b$) the Anderson model is defined by the Hamiltonian
\begin{eqnarray}
  H &=& \sum_{\vec k, \sigma } \epsilon_{\vec k} c_{\vec k \sigma}^+
  c_{\vec k \sigma} + \epsilon_d ({\sum_\sigma f_\sigma^+
  f_\sigma}+2 a^+a) +\nonumber\\
  &+& \sum_{\vec k,\sigma} \nonumber V_k (c_{\vec k \sigma}^+
  b^+ f_{\sigma}+ \sigma c_{\vec k \sigma}^+
  f_{-\sigma}^+ a + h.c.)  + U a^+a \nonumber;
\end{eqnarray}
where $\epsilon_d$ is the energy of the local $d$ level, $U$ is the
Coulomb interaction and $V_{\vec k}$ are the hybridization matrix
elements. A physical electron in the local level is created with the
operator $d_\sigma = b^+ f_\sigma + \sigma f_{-\sigma}^+ a$, where $b$
represents the empty level and $a$ the doubly occupied one (in the
following we will call $a$ and $b$ the heavy and the light boson,
respectively). For the representation to be exact, the local operator
constraint $Q\equiv \sum_{\sigma} f_\sigma^+ f_\sigma + b^+ b + a^+ a
= 1$ must be fulfilled at all times. This is guaranteed by a suitable
projection technique \cite{CTMA}.

\section{Method}
\begin{figure}[htb]
  \center{\epsfig{file=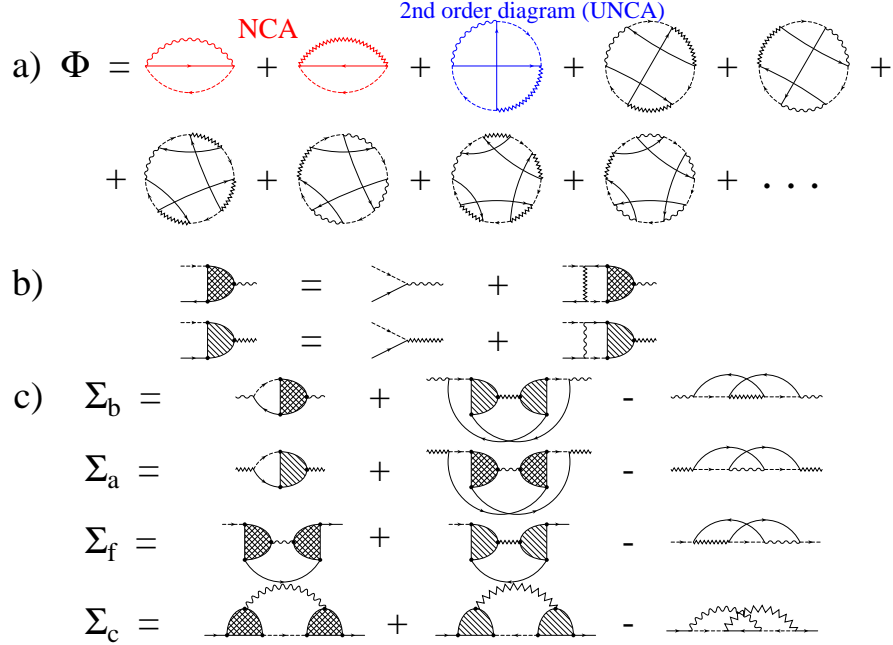,width=8.8cm,angle=-90}}
\caption{
  a)The Luttinger Ward functional, b) the Bethe-Salpeter equations for the
  T-matrix vertices (dark and light shaded areas), c) the Contributions to
  the irreducible self energies for pseudo-particles as well as for
  conduction electrons are shown.  }
\label{functional}
\end{figure}

It is well known that the Non-Crossing Approximation (NCA) gives the
correct Kondo temperature $T_K$ in the limit of large Coulomb
repulsion $U$ while it fails to recover the correct Kondo temperature
for finite U. The reason may be seen from the Schriffer-Wolff
transformation \cite{Schrieffer} where the effective coupling constant
has two equally important terms for $U\simeq -2\epsilon_d >0$
\begin{equation}
  J = -{V^2\over \epsilon_d} + {V^2 \over \epsilon_d + U}
\label{J}
\end{equation}
where just one of them survives in the large $U$ limit.  In
Eq.~(\ref{J}) the first (second) term is generated by exchange of a
bare light (heavy) boson. This implies that the light ($b$) and heavy
($a$) bosons should appear symmetrically in all bare diagrams. In order
to generate the corresponding diagrams of the Luttinger-Ward
functional one may start with the NCA diagram for the
$U\rightarrow\infty$ case (first diagram in Fig.~\ref{functional}a))
and replace the light boson line by a heavy boson, exchanging the
conduction electron lines appropriately. The result is the second
diagram. This is not sufficient, however, because the diagrams should
be symmetric in $a$, $b$ on the level of bare diagrams. One may show
that this exchange generates an infinite series of diagrams, each
containing a circle of pseudoparticle lines with one light boson and
$n$ heavy bosons (or vice versa) dressed by conduction electron lines
spanning at most three pseudoparticle lines (see
Fig.~\ref{functional}a)). The corresponding self-energies are obtained
by functional differentiation and are shown in Fig.~\ref{functional}b)
in terms of vertex functions (the dash- and cross-shaded half-circle
areas).

The Green's function of the physical $d$ electrons can be calculated
from the exact relation that can be derived with the help of the
equation of motion
\begin{equation}
G_{\vec k \vec{k^\prime}}(\omega) = \delta_{\vec k \vec{k^\prime}}
 G^0_{\vec k \vec k} (\omega) +{\sum }_{\vec k \vec k'} G_{\vec k
  \vec k}^0 (\omega)V_{\vec k}  G_d(\omega) V_{\vec{k^\prime}}^\ast
G_{\vec{k^\prime} \vec{k^\prime}}^0(\omega).
\label{EQOM}
\end{equation}
The local conduction electron Green's function can be expressed in
terms of the local conduction electron self energy $\Sigma_c$ (see
Fig.~\ref{functional}c)),
\begin{equation}
G_d(\omega) = {1\over V^2} \Sigma_c(\omega).
\label{Gd}
\end{equation}

\section{Numerical Results}

\begin{figure}[htb]
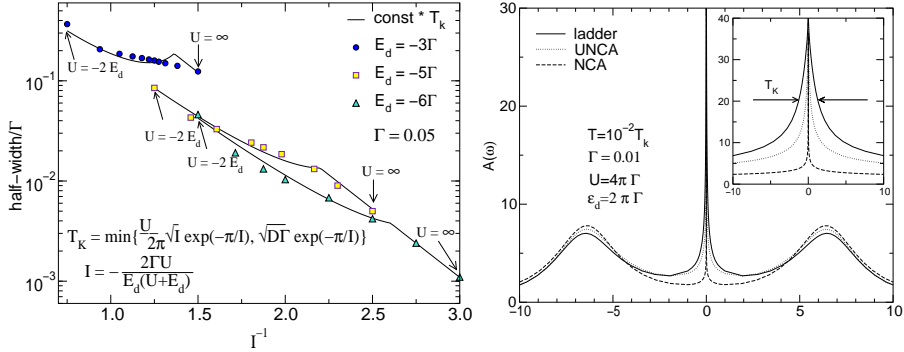

\begin{center}
  \epsfig{file=ImpTk2.eps,width=4.62cm,angle=-90}
  \epsfig{file=Costi1.eps,width=4.2cm,angle=-90}
\end{center}
\caption{
  Left: Kondo temperature for various parameters ($E_d$ and $U$).
  Solid lines are values of $T_k$ from the Bethe-ansatz solutions.
  Right: Local electron spectral function calculated with NCA, UNCA
  and the ladder approximation. The width of the peak is proportional to
  the Kondo temperature.}
\label{Kondo_temp}
\end{figure}

The Kondo temperature was extracted from the width of the
Abrikosov-Suhl peak in the local electron Green's function. The values we
get from the present approximation are in excellent agreement with the
exact values known from the Bethe ansatz solution, as can be seen in
Fig.~\ref{Kondo_temp}.  The simple Non-Crossing Approximation
(the first two terms in $\Phi$ in Fig.~\ref{functional}a) gives Kondo
temperatures several orders of magnitude too low. Including one additional
diagram does not increase $T_K$ sufficiently ($UNCA$) (here we
disagree with the corresponding claim in Ref. \cite{Pruschke}). An
{\it infinite} series of {\it skeleton} ladder diagrams is needed to get
the correct scaling of $T_K$ with the Coulomb energy $U$ and the local electron
energy $\epsilon_d$.

\section{Summary}
In summary, we have shown that an infinite number of skeleton diagrams
is necessary to recover the correct dynamic energy scale ($T_K$) in
the Anderson impurity model with finite Coulomb interaction energy.
We found a systematic way to obtain those diagrams and have shown that
the approximation is derivable a from Luttinger-Ward functional and
therefore conserving. Numerical results show excellent agreement with
the exact values of $T_K$.

  This work was funded by the ESF Program on {\it "Fermi-liquid
    instabilities in correlated metals"} (FERLIN), and by SFB 195 der
  Deutschen Forschungsgemeinschaft.


\begin{chapthebibliography}{99}
\bibitem{review} For an overview see {\it Mesoscopic Electron Transport}
L.~L.~Sohn, L.~P.~Kowenhoven, G.~Sch\"on, NATO ASI Series E, Vol. 345
(Kluwer, 1997).
\bibitem{Vollhardt} {A. Georges, G. Kotliar, W. Krauth, and
    M. J. Rozenberg, Rev. Mod. Phys. {\bf 68}, 13 (1996) }.
\bibitem{Schrieffer} {J. R. Schrieffer in P. A. Wolff, Phys. Rev. {\bf
    149}, 491 (1966)}.
\bibitem{CTMA} {
    J. Kroha, P. W\"olfle, and 
    T. A. Costi, Phys. Rev. Lett. {\bf 79}, 261 (1997);
    J. Kroha and P. W\"olfle, Acta Phys. Pol. B {\bf 29}, (12) 3781
    (1998), cond-mat/9811074;
    J. Kroha and P. W\"olfle, Andvances in Solid State Phys. {\bf 39}, 271 (1999)}.
\bibitem{Pruschke} {Pruschke et al. Z. Phys. B {\bf 74}, 439-449 (1989)}.
\end{chapthebibliography}
\end{document}